\documentclass[acmsmall]{acmart}
\setcopyright{none}

\AtBeginDocument{%
  \providecommand\BibTeX{{%
      \normalfont{} \kern-0.5em{\scshape i\kern-0.25em b}\kern-0.8em\TeX}}}

\usepackage[T1]{fontenc}
\usepackage{adjustbox}
\usepackage[flushleft]{threeparttable}
\usepackage{booktabs} 
\usepackage{multirow}
\usepackage{enumitem}
\usepackage{array}
\usepackage{siunitx}
\usepackage{soul}
\usepackage{mathtools,bm,amsmath,amsfonts}
\usepackage{nicefrac}
\usepackage{url}
\usepackage{graphicx,subcaption}
\usepackage{cleveref}
\usepackage{tikz}
\usetikzlibrary{arrows, decorations.markings}

\tikzstyle{vecArrow} = [thick, decoration={markings,mark=at position
   1 with {\arrow[semithick]{open triangle 60}}},
   double distance=1.4pt, shorten >= 5.5pt,
   preaction = {decorate},
   postaction = {draw,line width=1.4pt, white,shorten >= 4.5pt}]
\tikzstyle{innerWhite} = [semithick, white,line width=1.4pt, shorten >= 4.5pt]
\makeatletter
\pgfdeclareshape{document}{
\inheritsavedanchors[from=rectangle] 
\inheritanchorborder[from=rectangle]
\inheritanchor[from=rectangle]{center}
\inheritanchor[from=rectangle]{north}
\inheritanchor[from=rectangle]{south}
\inheritanchor[from=rectangle]{west}
\inheritanchor[from=rectangle]{east}
\backgroundpath{
\southwest \pgf@xa=\pgf@x \pgf@ya=\pgf@y
\northeast \pgf@xb=\pgf@x \pgf@yb=\pgf@y
\pgf@xc=\pgf@xb \advance\pgf@xc by-10pt 
\pgf@yc=\pgf@yb \advance\pgf@yc by-10pt
\pgfpathmoveto{\pgfpoint{\pgf@xa}{\pgf@ya}}
\pgfpathlineto{\pgfpoint{\pgf@xa}{\pgf@yb}}
\pgfpathlineto{\pgfpoint{\pgf@xc}{\pgf@yb}}
\pgfpathlineto{\pgfpoint{\pgf@xb}{\pgf@yc}}
\pgfpathlineto{\pgfpoint{\pgf@xb}{\pgf@ya}}
\pgfpathclose
\pgfpathmoveto{\pgfpoint{\pgf@xc}{\pgf@yb}}
\pgfpathlineto{\pgfpoint{\pgf@xc}{\pgf@yc}}
\pgfpathlineto{\pgfpoint{\pgf@xb}{\pgf@yc}}
\pgfpathlineto{\pgfpoint{\pgf@xc}{\pgf@yc}}
}
}
\makeatother

\usepackage[T1]{fontenc}
\crefname{figure}{Figure}{Figures}

\usepackage{listings}
\definecolor{mygreen}{HTML}{859900}
\definecolor{myblue}{HTML}{268bd2}
\definecolor{myred}{HTML}{cb4b16}
\definecolor{mypurple}{HTML}{6c71c4}

\begin{document}


\title{How Flexible is Your Computing System?}
\author{Shihua Huang}
\email{shihuahuang94@gmail.com}
\affiliation{\institution{Prodrive Technologies}}
\authornote{These authors contributed equally}
\author{Luc Waeijen}
\affiliation{\institution{GrAI Matter Labs}}
\email{lwaeijen@graimatterlabs.ai}
\authornotemark[1]
\author{Henk Corporaal}
\affiliation{\institution{Eindhoven University of Technology}}
\email{h.corporaal@tue.nl}

\renewcommand{\shortauthors}{Huang, Waeijen, and Corporaal}

\begin{abstract}
In literature computer architectures are frequently claimed to be \textsl{highly flexible}, typically implying there exist trade-offs between flexibility and performance or energy efficiency.
Processor flexibility, however, is not very sharply defined, and as such these claims can not be validated, nor can such hypothetical relations be fully understood and exploited in the design of computing systems.
This paper is an attempt to introduce scientific rigour to the notion of flexibility in computing systems.
A survey is conducted to compile an overview of references to flexibility in literature both in the computer architecture domain as well as related fields.
A classification is introduced to categorize different views on flexibility, which ultimately form the foundation for a qualitative definition of flexibility.
Departing from the qualitative definition of flexibility, a generic quantifiable metric is proposed, enabling valid quantitative comparison of the flexibility of various architectures.
To validate the proposed method, and evaluate the relation between the proposed metric and the general notion of flexibility, the flexibility metric is measured for 25 computing systems, including CPUs, GPUs, DSPs, and FPGAs, and 40 ASIPs taken from literature.
The obtained results provide insights into some of the speculative trade-offs between flexibility and properties such as energy efficiency and area efficiency.
Overall the proposed quantitative flexibility metric shows to be commensurate with some generally accepted qualitative notions of flexibility collected in the survey, although some surprising discrepancies can also be observed.
The proposed metric and the obtained results are placed into context of the state of the art on compute flexibility, and extensive reflection provides not only a complete overview of the field, but also discusses possible alternative approaches and open issues.
Note that this work does not aim to provide a final answer to the definition of flexibility, but rather provides a framework to initiate a broader discussion in the computer architecture society on defining, understanding, and ultimately taking advantage of flexibility.

\end{abstract}



\maketitle

\section{Introduction}
Arguably one of the most famous books in the field is ``Computer Architecture --- A Quantitative approach'' by John L. Hennessy and David A. Patterson~\cite{hennessypatterson}.
The title itself is concise and apt, so it is interesting the authors opted to add this particular subtitle: a \textsl{quantitative} approach.
It implies the belief that quantifying design choices ultimately leads to better computer architectures, a message that certainly could be directed towards those who make claims about \textsl{flexible} architectures without means of quantifying these claims, or even without as much as a commonly accepted qualitative definition of flexibility.
With Moore's law seemingly coming to an end, new advancements in computing will have to be made on the architectural side.
To advance the state of the art, fundamental understanding of various trade-offs in computer design is vital.
The way forward therefore, is a quantitative one.

Many key system properties such as performance, power dissipation, and energy efficiency are all well defined in a quantitative manner.
With these metrics in place, quantitative, objective comparisons can be conducted between different machines.
For flexibility however, such a quantitative (and even qualitative!) definition is lacking, despite it's increasing importance in system design.
In product research and development, computing platforms are required to sufficiently support new or updated algorithms, as algorithms are changing at a striking speed.
Exemplary are the current developments in artificial intelligence, which result in new compute-intensive algorithms at a high cadence.
Such rapidly developing markets require systems that can deal with changing applications, which is the property flexibility typically seems to refer to.
However, in absence of a proper definition, it is impossible to make solid statements, and compare designs on flexibility.

\begin{figure}[h!]
\captionsetup[subfigure]{position=b}
\centering
    \hfill%
    \subcaptionbox{%
        Flexibility related to performance and power according to Tobias Noll~\cite{Tnoll,dashen}.
        Note that according to this figure flexibility is directly related to power dissipation.%
        \label{fig:intro_ref}
    }{
        \includegraphics[width=0.35\linewidth]{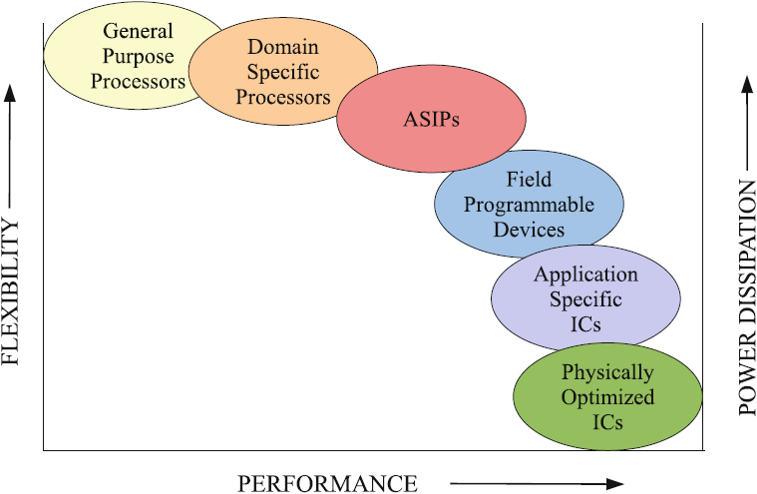}
        \Description[]{Graph with on the horizontal axis performance, left vertical axis flexibility, and right vertical axis power dissipation. Architectures are grouped, roughly following a parabolic function with it's top flexibility/power dissipation at the lowest performance point, sloping down towards zero as performance increases. For left to right the architecture groups are: General purpose processors, domain specific processors, ASIPs, field programmable devices, application specific ICs, physically optimized ICs}
    }%
    \hfill%
    \subcaptionbox{%
        Flexibility versus performance plot by Geoffrey Ndu~\cite{randomflex1}.%
        The ordering of the flexibility of architecture classes aligns quite well with the ordering given in Figure~\ref{fig:intro_ref}, although curve is quite different.%
        \label{fig:intro_ref1}
    }{
        \includegraphics[width=0.35\linewidth]{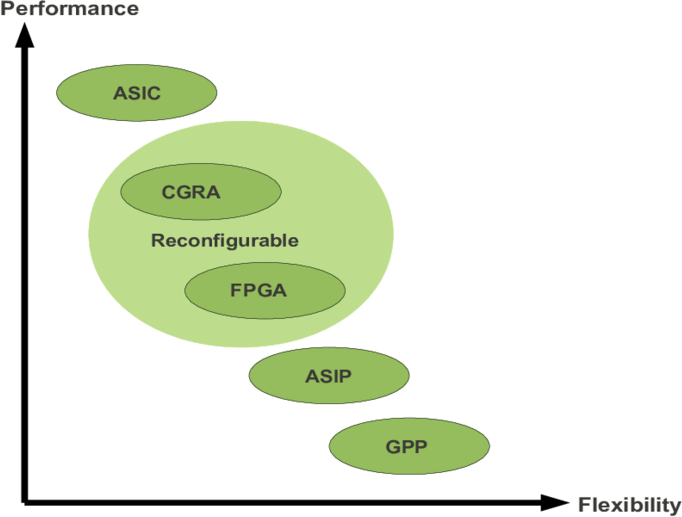}
    }\hfill\hspace{1em}

    \hfill%
    \subcaptionbox{%
        `Application Flexibility` versus `Performance Efficiency` by Markus Willems from Synopsys~\cite{randomflex4}.
        Unfortunately no further definition of both these metrics is provided in the accompanying text.%
        \label{fig:intro_ref4}
    }{
        \includegraphics[width=0.35\linewidth]{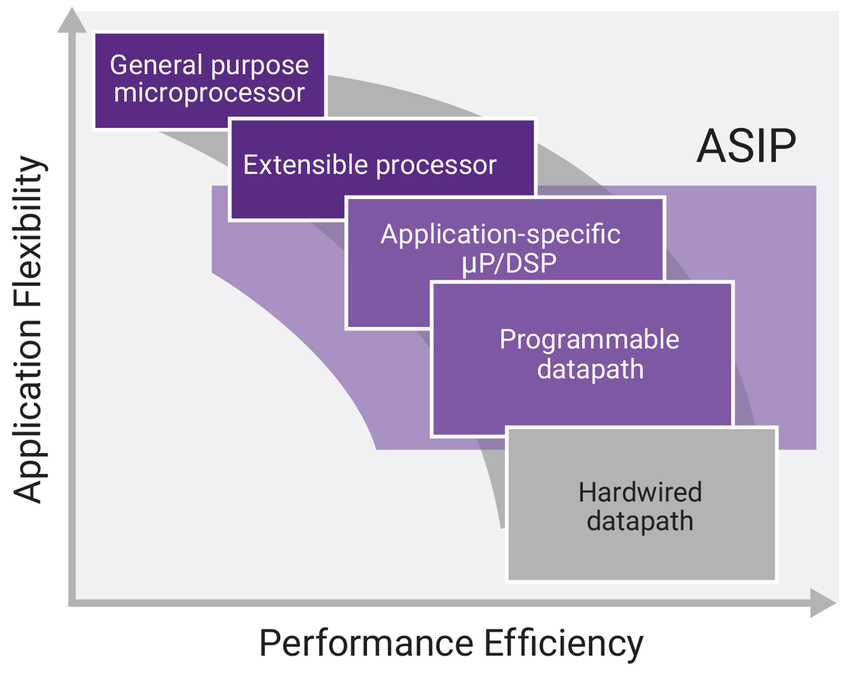}
    }\hfill%
    \subcaptionbox{%
        An interesting graph by Ahmed Osman El-Rayis~\cite{randomflex5} which relates many metrics including flexibility.%
        In contrast to Figure~\ref{fig:intro_ref} flexibility is here claimed to be directly related to area.
        \label{fig:intro_ref5}
    }{
        \includegraphics[width=0.35\linewidth]{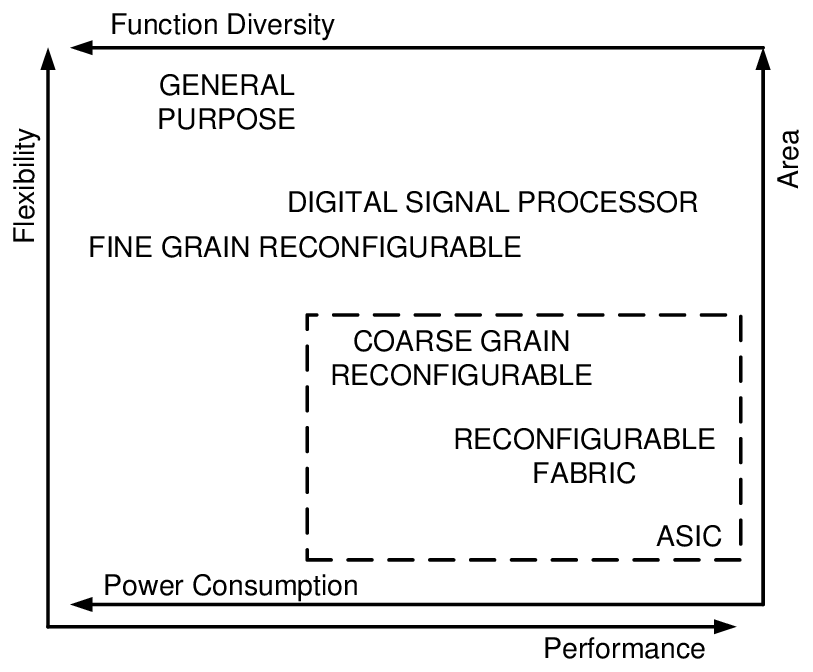}
    }%
    \hfill\hspace{1em}
\caption{Collection of published figures with claims about the flexibility of architecture classes, and relations between flexibility and other metrics such as performance and power. Note that none of the axis in these figures are labelled with units.}%
\label{fig:intro_ref_base}
\end{figure}

Despite the lack of a formal definition of flexibility, there appear to be some commonly accepted notions surrounding flexibility.
In particular, flexibility seems mainly used to refer to the adaptability of processors to different applications.
This leads to the common idea that a programmable processor which can be reused across applications is `flexible'.
On the other hand, a processor with fixed logic such as an ASIC cannot adapt, exposing its inflexibility~\cite{flex_def}.
As can be seen in Figure~\ref{fig:intro_ref_base} the authors of these figures appear to agree with this sentiment.
However there are also some contradictions to this view on flexibility however.
For example in Figure~\ref{fig:intro_ref}, among programmable processors the field-programmable devices are claimed by the authors to be less flexible than software programmable processors due to their inadequate programmability~\cite{dashen}.
Unfortunately the term ``programmability'' is also ill-defined here.
Perhaps the best definition of programmability in existence is to check Turing completeness of a programmable device, but this would leave only two classes of programmability making it a measure with low practical value.
Another perspective on flexibility refers to how well a processor supports different applications, in which case FPGAs could be seen as the most flexible, since any hardware, including DSPs, GPUs, and CPUs, can be instantiated on FPGAs.
Apart from this debate on how to rank the flexibility of architecture classes, perhaps even more worrisome are the contradicting claims on relations between flexibility and other metrics.
In Figure~\ref{fig:intro_ref5} Ahmed Osman El-Rayis equates flexibility to area, whereas Tobias Noll sees it as directly related to power dissipation in Figure~\ref{fig:intro_ref}.
While this is definitely not an exhaustive list of views on flexibility, it painfully exposes how the lack of a formal definition leads to a wild-west of claims and conflicting visions, none of which can be backed up with objective measurements.

Despite the greatly varying interpretations of flexibility, many seem to agree that there may be interesting relations and trade-offs between flexibility and other properties, such as performance and energy efficiency.
As illustrated in Figure~\ref{fig:intro_ref_base}, processing platform alternatives have been evaluated and ranked in terms of flexibility, performance, power dissipation, and area~\cite{Tnoll,dashen,randomflex1,randomflex4,randomflex5}.
A variety of architectures have been developed which claim to balance energy efficiency and flexibility.
The development of domain-specific functional units and a transition to heterogeneous multi-core systems are a testament to this notion~\cite{hetero, phdguy, Energy-Efficient}.
These hypothetical relations suggest that understanding flexibility is beneficial when designing a system, enabling informed trade-offs.

To overcome the lack of understanding of flexibility, this work sets out to provide both a qualitative and quantitative definition of flexibility.
It should be noted though that, with such a fragmented landscape of interpretations of flexibility, the authors are under no illusion that it is possible to unify the field and reach consensus without a wider discussion.
Instead, this work is to be seen as a first attempt, which does not so much aspire to provide a definitive answer, as it hopes to be thought provoking and spark a discussion within the community.

To arrive at a quantitative measure for flexibility, a qualitative definition is established by exploring uses of the term flexibility in literature and then examining various options.
Based on this qualitative definition, a quantitative measure is derived.
In the translation of flexibility, from a qualitative term to a quantitative definition, there exist several degrees of freedom.
The final choices made in this translation are motivated extensively.
However, more importantly, the alternatives are discussed systematically in similar detail.
The intention is that this systematic approach can provide an initial framework for a broader discussion in the community on how flexibility should be defined, such that eventually a standard accepted metric can be determined.

To validate the metric proposed in this work, in total 14 applications are benchmarked on 25 different commercial off the shelf (COTS) platforms.
It is shown that results align with several common concepts of flexibility found in the literature.
For example, GPUs deliver the highest performance in general for the used parallel benchmarks but sacrifice in terms of flexibility, compared to general purpose processors (GPPs) in Figure~\ref{fig:intro_ref}.
Furthermore the flexibility of 40 application specific architectures from literature is determined, to evaluate the relation between specialisation and flexibility.

The main contributions of this work can be summarized as follows:
\begin{enumerate}
    \item Survey of flexibility in computing and other engineering fields;
    \item A qualitative and quantitative definition of flexibility;
    \item Definition of intrinsic workload to normalize performance, and accompanying open-source tool~\cite{workloadestimator} for automated extraction;
    \item Evaluation of the proposed metric on 25 COTS platforms for 14 benchmarks, and 40 application specific processors taken from literature;
    \item In-depth Comparison with alternative definitions of flexibility/versatility;
    \item Extensive discussion on the proposed metric, and various alternatives, placing it in context of the current state of the art.
\end{enumerate}

\textbf{NOTE: This document is a partial preprint. Full text will be made available after publication elsewhere.}

\bibliographystyle{ACM-Reference-Format}
\bibliography{bibliography}

\end{document}